\documentclass[conference,10pt]{IEEEtran}
\IEEEoverridecommandlockouts
\usepackage{cite}
\usepackage{amsmath,amssymb,amsfonts}

\newtheorem{remark}{Remark}
\usepackage{graphicx}
\usepackage{textcomp}
\usepackage{xcolor}
\usepackage{multirow}
\usepackage{threeparttable}
\def\BibTeX{{\rm B\kern-.05em{\sc i\kern-.025em b}\kern-.08em
    T\kern-.1667em\lower.7ex\hbox{E}\kern-.125emX}}

\newtheorem{example}{Example}

\newcommand{\blue}[1]{#1}

\usepackage[normalem]{ulem}

\newcommand{\tket}{t$\ket{\mathrm{ket}}$}
\newcommand{\swap}[2]{\mathrm{SWAP}(#1,#2)}

\newcommand{\ket}[1]{\left| {#1} \right\rangle }
\newcommand{\dist}{\textsf{dist}}
\newcommand{\cost}[1]{\textsc{cost}(#1)}
\newcommand{\astate}{{\mathbf{s}}}

\usepackage[ruled,vlined]{algorithm2e}
\SetKwRepeat{Do}{do}{for}

\begin{document}

\title{
Supervised Learning Enhanced Quantum Circuit Transformation
}

\author{\IEEEauthorblockN{1\textsuperscript{st} Xiangzhen Zhou}
\IEEEauthorblockA{
\textit{Centre for Quantum Software and Information} \\
\textit{University of Technology Sydney}\\
Australia \\
xiangzhen.zhou@uts.edu.au}
\and
\IEEEauthorblockN{2\textsuperscript{nd} Yuan Feng}
\IEEEauthorblockA{
\textit{Centre for Quantum Software and Information} \\
\textit{University of Technology Sydney}\\
Australia \\
yuan.feng@uts.edu.au}
\and
\IEEEauthorblockN{3\textsuperscript{rd} Sanjiang Li}
\IEEEauthorblockA{
\textit{Centre for Quantum Software and Information} \\
\textit{University of Technology Sydney}\\
Australia \\
sanjiang.li@uts.edu.au}
}

\author{Xiangzhen Zhou\thanks{Xiangzhen Zhou is with State Key Lab of Millimeter Waves, Southeast University, Nanjing 211189, China and Centre for Quantum Software and Information, Faculty of Engineering and Information Technology, University of Technology Sydney, NSW 2007, Australia.}, Yuan Feng\thanks{Yuan Feng is with Centre for Quantum Software and Information, Faculty of Engineering and Information Technology, University of Technology Sydney, NSW 2007, Australia (e-mail: yuan.feng@uts.edu.au).}, and Sanjiang Li\thanks{Sanjiang Li is with Centre for Quantum Software and Information, Faculty of Engineering and Information Technology, University of Technology Sydney, NSW 2007, Australia (e-mail: sanjiang.li@uts.edu.au).}}

\maketitle

\begin{abstract}
\blue{A q}uantum circuit transformation (QCT) is required when executing a quantum program in a real quantum processing unit (QPU). Through inserting auxiliary SWAP gates, a QCT algorithm transforms a quantum circuit to one that satisfies the connectivity constraint imposed by the QPU. Due to the non-negligible gate error and the limited qubit coherence time of the QPU, QCT algorithms which minimize gate number or circuit depth or maximize the fidelity of output circuits are in urgent need. Unfortunately, finding optimized transformations often involves exhaustive search\blue{es}, which \blue{are} extremely time-consuming and not practical for \blue{most} circuits. In this paper, we propose a framework \blue{that} uses a policy artificial neural network (ANN) trained by supervised learning on shallow circuits to help existing QCT algorithms select the most promising SWAP \blue{gate}. ANNs can be trained \blue{off-line} in a distributed way. The trained ANN can be easily incorporated into QCT algorithms without bringing too much overhead in time complexity.
Exemplary embeddings of the trained ANNs into target QCT algorithms demonstrate that the transformation performance can be consistently improved on QPUs with various connectivity structures and random \blue{or realistic} quantum circuits.
\end{abstract}

\begin{IEEEkeywords}
quantum circuit transformation, machine learning, supervised learning, artificial neural network
\end{IEEEkeywords}

\section{Introduction}
It is widely recognized that Moore's law, which states that the number of transistors in a dense integrated circuit doubles about every two years, will, if not already \blue{has}, come to an end in the near future. On the other hand, although still in their infancy, quantum computers or, more precisely, quantum processing units (QPUs) have seen a steady increase 
in the number of valid qubits in the past several years. QPUs in the Noisy Intermediate-Scale Quantum (NISQ) era have rather limited qubit coherence time and only support a few kinds of one- or two-qubit elementary quantum gates, which usually have non-negligible gate error. Nevertheless, quantum supremacy was demonstrated in Sycamore, Google's recent 53-qubit QPU \cite{HarrowSupremacy}. More and more quantum or hybrid quantum-classical algorithms have been designed for these NISQ era QPUs \cite{PreskillNISQ}.
Naturally, the size \blue{(i.e., number of gates)} and depth of such a quantum algorithm (or, a quantum circuit) are limited, due to the error caused by the decoherence and noise inherently present in these QPUs. Moreover, current QPUs impose strict connectivity constraints which require that any two-qubit operation can only be applied between connected qubits. This presents a challenge for quantum computing in the NISQ era. Assume that all quantum gates in a quantum circuit $C$ have already been decomposed into elementary gates supported by the QPU. Before executing $C$, we need to  transform $C$ into a functionally equivalent one while obeying the connectivity constraints imposed by the QPU. This process was first considered in \cite{MaslovFM07} and has many different names. In this paper, following \cite{Childs}, we term it as \emph{quantum circuit transformation} (QCT).  

Usually, the QCT process will introduce a large number of auxiliary SWAP gates, which will, in turn, significantly \blue{decrease the fidelity of the output}. \blue{Therefore, algorithms need to be designed that} can minimize \blue{the} gate number or circuit depth and/or maximize the fidelity of the circuit \cite{wille2018computer}. While it is not difficult to transform a circuit into a functionally equivalent one that satisfies the connectivity constraints, the real challenge lies in finding an optimal one. Currently, there are a few exact QCT algorithms which can construct an equivalent executable circuit with either the minimal number of auxiliary SWAPs \cite{LyeMinSwap,SAT} or the smallest circuit depth \cite{tan2020optimal}. The problem with these exact algorithms is that they are extremely time-consuming and can only process quantum circuits with very small size and very shallow depth on QPUs with very small number of qubits. \blue{For example, it was shown in \cite{SAT} that the exact solution can be computed within an acceptable time only for circuits with no more than 5 qubits and 100 gates.} As a consequence, most existing algorithms are approximate. Roughly speaking, these approximate algorithms can be further divided into two categories. Algorithms in the first category reformulate the QCT problem and apply some off-the-shelf tools to solve it \cite{BoothPlanning,VenturelliPlanning}, while those in the second use heuristic search to construct the output circuit step-by-step \cite{Astar,LiUCSB,CowtanRouting,SAHS,FiDLS,initialmap,SiraichiAllocation,FiniganAllocationNISQ}. As empirically evaluated in \cite{tan2020optimality}, these algorithms are still very far from being optimal. Take the industry-level QCT algorithm developed by the Cambridge Quantum Computing (addressed as \blue{\tket} henceforth) as an example. It was shown in \cite{tan2020optimality} that, for IBM Q Tokyo and input circuits with depths from 5 to 45, the optimality gap (the ratio of the output circuit depth of \blue{\tket} to the optimal depth) could still be as high as 5x! Meanwhile, it 
is worth mentioning that there are QCT algorithms that have  significantly better outputs than \blue{\tket}. The Monte Carol Tree Search (MCTS) based algorithm devised in \cite{MCTS_QCT}, called MCTS in this paper, seems to be the best reported QCT algorithm on IBM Q Tokyo, which inserts in average 60\% less SWAP gates than \blue{\tket} on a set of 114 real benchmark circuits. 

Inspired by the recent success of \emph{artificial neural network} (ANN) \cite{ANN} in enhancing the MCTS algorithm adopted by AlphaGo \cite{alphago1}, we propose a framework in which a policy ANN is trained by supervised learning on shallow circuits to help existing QCT heuristic search algorithms select the most promising SWAP. Supervised learning \cite{ML} is the machine learning paradigm of learning a function that maps an input to an output under the supervision of a `teacher' whose role is to provide a set of training examples, represented as a set of labeled training data. For each connectivity structure, such a policy ANN could be trained by using an (almost) exact algorithm or the target QCT algorithm. This is very attractive as ANNs can be trained
in a distributed way off-line
and more precise training data can be obtained and accumulated by applying (time-consuming) exact or near-exact QCT algorithms on shallow random circuits. Moreover, the trained policy ANN can be embedded in the target heuristic search-based QCT algorithm to enhance its performance. We provide two exemplary embeddings, one uses the SAHS algorithm \cite{SAHS} and the second the MCTS algorithm \cite{MCTS_QCT} \blue{(cf. Sec.s~\ref{sec:sahs}~and~\ref{sec:mcts} for their detailed implementations)}. Empirical results on QPUs with various connectivity structures and random \blue{or realistic} quantum circuits demonstrate that the performance of both \blue{SAHS} and MCTS can be consistently improved by employing the trained policy ANNs.

In the literature, there are also several QCT algorithms which have exploited machine learning techniques. In \cite{ML_parameter}, machine learning is used to optimize the hyper-parameters of QCT algorithms, not being directly involved in the transformation process.
Reinforcement learning is utilized in 
\cite{RL2} to reduce the depth of the transformed quantum circuit.
Different from these works, the proposed policy ANN can be embedded in many existing search-based QCT algorithms to enhance their performance, and the experimental results in Sec.~\ref{sec:sahs_exp} and \ref{sec:mcts_exp} demonstrate that the improvement is obvious and consistent.

The remainder of this paper is organized as follows. After a brief introduction of the QCT problem in Sec.~\ref{sec:qct}, we describe in detail the modules of the proposed framework and validate the efficacy of the trained ANN and its embedding process in Sec.~\ref{sec:framework}. Two exemplary applications of the proposed framework based on different state-of-the-art QCT algorithms are then introduced in Sec.~\ref{sec:sahs} and \ref{sec:mcts}. The last section concludes the paper with an outlook for future research. \blue{The scalability of our framework in terms of the qubit number is discussed in Appendix.}

\section{The Quantum Circuit Transformation Problem}\label{sec:qct}
In classical computing, binary digit, or \emph{bit}, is the basic unit of information which has only two states, 0 or 1. In contrast, quantum bit, or \emph{qubit}, serves as the basic unit of quantum information, which can be in the superposition $\alpha \ket{0} + \beta \ket{1}$ of the two basis states, denoted $\ket{0}$ and $\ket{1}$ respectively, where $\alpha,\beta$ are  complex numbers and ${\left| \alpha  \right|^2} + {\left| \beta  \right|^2} = 1$.

\begin{figure}[tbp]
\centerline{\includegraphics[width=0.3\textwidth]{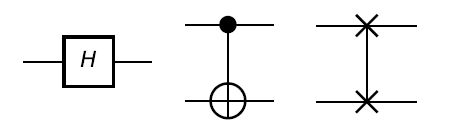}}
\caption{Hadamard, CNOT and SWAP gates (from left to right).}
\label{fig:gates}
\end{figure}

\begin{figure}[tbp]
\centerline{\includegraphics[width=0.46\textwidth]{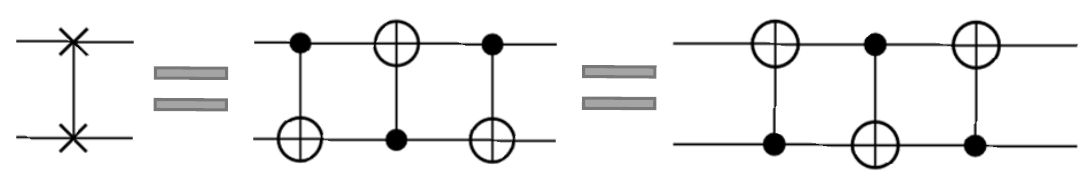}}
\caption{The decomposition of a SWAP into three CNOT gates.}
\label{fig:swap_decom}
\end{figure}

\begin{figure}[tbp]
	\centerline{\includegraphics[width=0.46\textwidth]{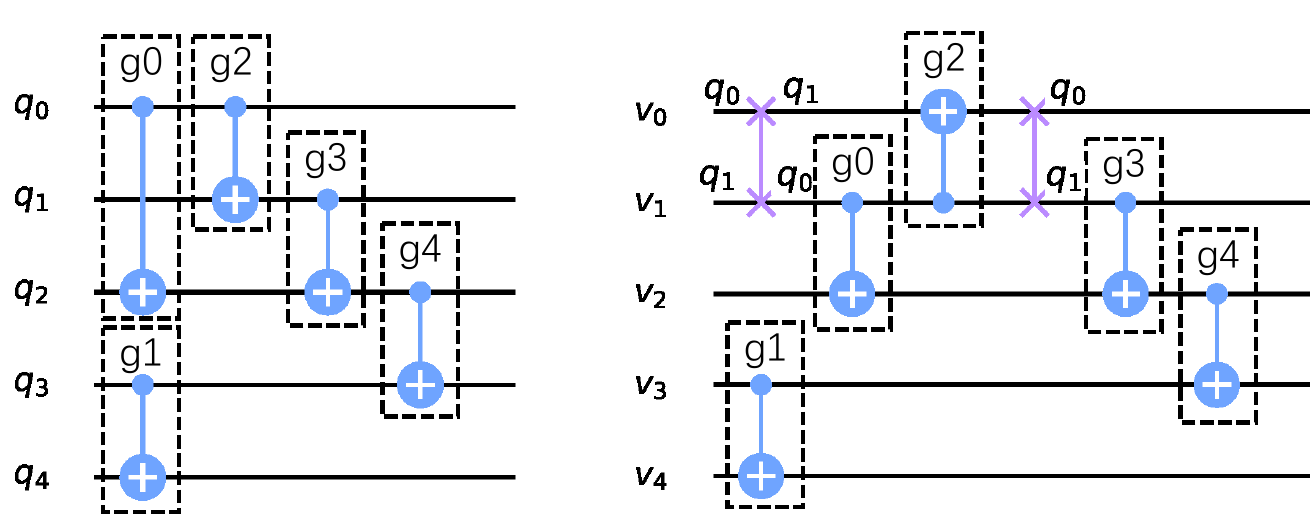}}
	\caption{A quantum circuit (left) and a functionally equivalent  circuit (right), which is executable on IBM Q Tokyo and obtained by inserting two SWAPs and starting with the naive initial mapping that maps $q_i$ to $v_i$ for $0\leq i\leq 4$.}
	\label{fig:fig_cir}
\end{figure}

States of qubits can be manipulated by \emph{quantum gates}. Depicted in Fig.~\ref{fig:gates} are graphic representation of three quantum gates: Hadamard, CNOT and SWAP. Hadamard is a single-qubit gate that has the ability to generate superposition: it maps $\ket{0}$ to $(\ket{0}+\ket{1})/\sqrt{2}$ and $\ket{1}$ to $(\ket{0}-\ket{1})/\sqrt{2}$. CNOT and SWAP are both two-qubit gates. CNOT flips the target qubit depending on the state of the control qubit; that is, CNOT: $\ket{c}\ket{t} \rightarrow \ket{c}\ket{c\oplus t}$, where $c,t\in \{0,1\}$ and $\oplus$ denotes exclusive-or. SWAP exchanges the states of its operand qubits: it maps $\ket{a}\ket{b}$ to $\ket{b}\ket{a}$ for all $a,b\in \{0,1\}$.
Note that, as shown in Fig.~\ref{fig:swap_decom}, one SWAP gate can be decomposed into three CNOT gates.

Quantum algorithms are often expressed as \emph{quantum circuits} each \blue{of which consists} of a set of qubits and a sequence of quantum gates. Shown in Fig.~\ref{fig:fig_cir} (left) is a quantum circuit with 5 qubits and 5 gates.
The gates in a quantum circuit can be divided into different \emph{layers} such that gates in the same layer can be executed simultaneously. The first or front layer of circuit $C$ is  denoted by $\mathcal{L}_0(C)$. Likewise, for any $i\geq 1$, $\mathcal{L}_{i-1}(C)$ represents the $i$-th layer of $C$.

\begin{figure}[tbp]
	\centerline{\includegraphics[width=0.44\textwidth]{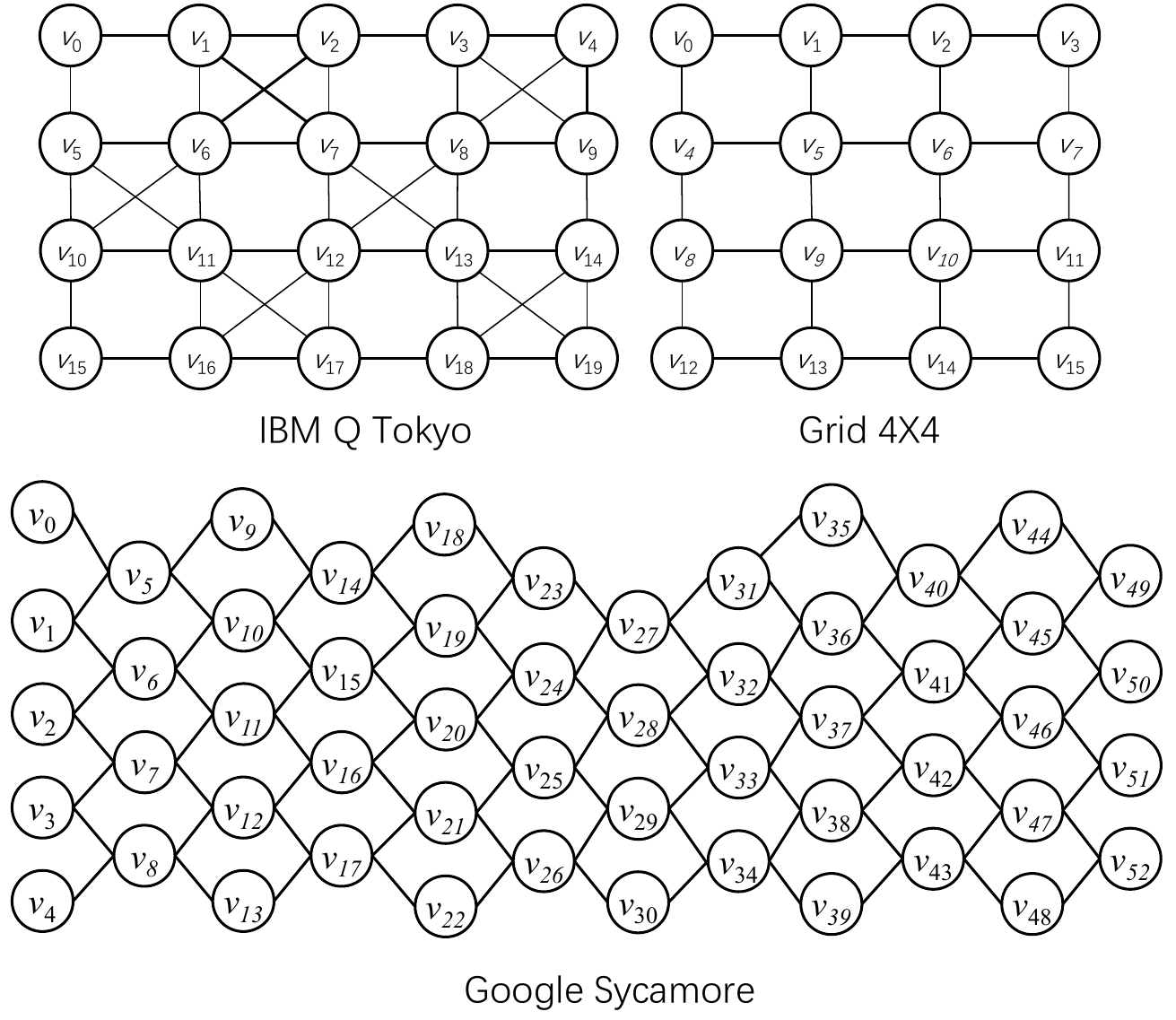}}
	\caption{The architecture graphs for IBM Q Tokyo, Grid 4X4 and Google Sycamore.}
	\label{fig:AG}
\end{figure}

In a QPU, only a limited set of quantum gates, called \emph{elementary gates}, can be directly executed. Without loss of generality, we assume that the elementary gates of a QPU form a universal set of quantum gates, which consists of the (two-qubit) CNOT gate and some single-qubit gates. Furthermore, we represent the connectivity structure of the QPU as an undirected graph, $AG = \left( {V,E} \right)$, called the \emph{architecture graph} \cite{Childs} (see Fig.~\ref{fig:AG} for a few examples), in which each node represents a qubit and two qubits are connected if and only if  CNOT gates can be applied between them. 

Before executing a quantum circuit on a QPU, two procedures need to be done. The first one is to decompose the gates in the circuit into elementary gates \cite{ShendeSynthesis} and obtain an equivalent one which we call the elementary circuit; the second is to transform the elementary circuit into one that satisfies the connectivity constraints imposed by the QPU while not changing its functionality. This latter procedure is called \emph{quantum circuit transformation} (QCT) \cite{Childs}. Henceforth, we will call the input elementary quantum circuit of QCT \emph{logical circuit}, its qubits \emph{logical qubits}, the output circuit \emph{physical circuit}, and its qubits \emph{physical qubits}. 
In this paper, we only consider the 
QCT procedure. Furthermore, as single-qubit elementary gates can be directly executed on a QPU, we assume that the logical circuits to be transformed consist solely of CNOT gates.

To transform a logical circuit $LC$ to a physical one executable on a QPU, we first map (or, allocate) the logical qubits in $LC$ to the physical qubits in $V$.
A two-qubit (CNOT) gate in the front layer of $LC$ is \emph{executable} if the allocated physical qubits of its operand logical qubits are adjacent in the architecture graph $AG$ of the QPU.
Note that in general it is unlikely that all gates in $LC$ are executable by a single mapping. Once no gates are executable by the current mapping $\tau$, a QCT algorithm seeks to insert into the circuit one or more ancillary SWAP gates to change $\tau$ into a new mapping so that more gates are executable. This insertion-execution process is iterated until all gates from $LC$ are executed. Fig.~\ref{fig:fig_cir} (right) shows a physical circuit transformed from the logical circuit on the left. 

The objective of the QCT procedure may vary in different algorithms, e.g., gate count \cite{Astar,LiUCSB}, depth \cite{tan2020optimal} and fidelity \cite{Rodney}. In this paper, we only consider algorithms which aim to minimize the total number of CNOT gates in the output physical circuit. Recall that  each SWAP can be decomposed into 3 CNOT gates as shown in Fig.~\ref{fig:swap_decom}. This is equivalent to minimizing the number of inserted SWAP gates.

\section{Supervised Learning for Quantum Circuit Transformation}
\label{sec:framework}

Recall that the main idea behind the QCT process is to insert SWAP gates step-by-step to construct the physical circuit.
\blue{Hence, the strategy used to select the most promising SWAP among the candidate set often has a significant impact on the performance of the QCT algorithm.}
A wide range of QCT algorithms utilize heuristic-based evaluation functions to assist this process. Whereas, this evaluation strategy is usually `short-sighted' and only able to take the information in the current state into consideration. To tackle this issue, a trained policy ANN can be used to boost the accuracy of the evaluation process. 


\blue{The idea of our ANN-based framework is 
to first train an ANN using a `feeding' QCT algorithm, say QCT-A, and then boost the  target QCT algorithm (possibly different from QCT-A), say QCT-B, with the trained ANN. In this section, after describing the ANN-based framework in detail, we show that the ANN trained itself can be directly used for quantum circuit transformation. 
Furthermore, we introduce a baseline QCT algorithm, called BASE, and then demonstrate how to boost BASE with a trained ANN. 
}

\blue{For all experimental evaluations in this paper, we use Python as the programming language, and all experiments are done on a laptop with i7-11800 CPU, 32 GB memory and RTX 3060 GPU. We use both random and realistic circuits as benchmarks, which, together with detailed experimental data, can be found in GitHub\footnote{https://github.com/BensonZhou1991/Supervised-Learning-Enhanced-Quantum-Circuit-Transformation/}.}

\subsection{Details of the Framework}
\label{sec:details}
\begin{figure*}[t]
	\centerline{\includegraphics[width=0.75\textwidth]{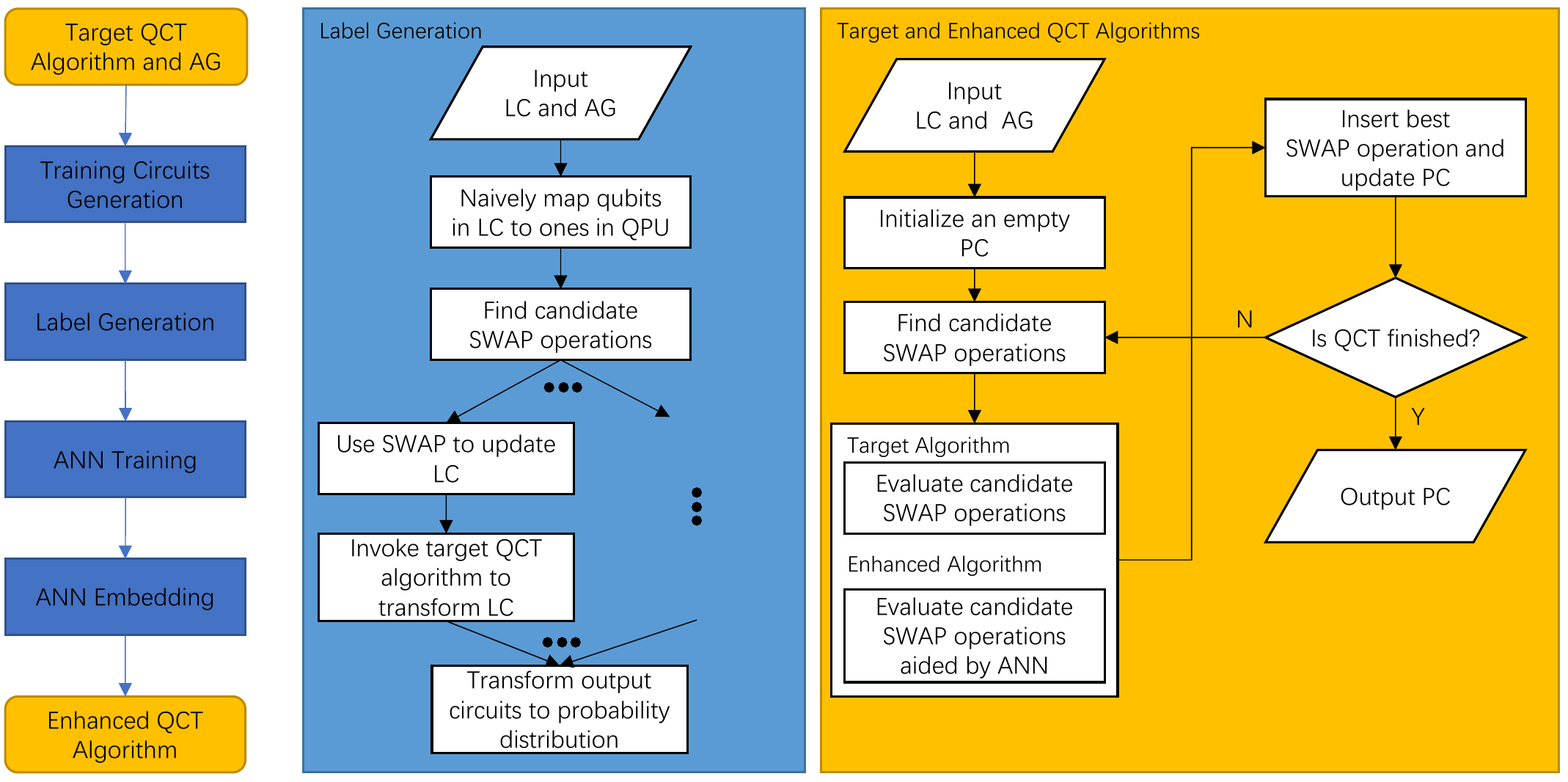}}
	\caption{The modules composing the proposed framework (left) and the system diagrams for Label Generation (middle) and (ANN-embedded) QCT algorithm (right), \blue{where LC and AG stand for logical circuit and architecture graph, respectively.}
	}
	\label{fig:framework}
\end{figure*}

\blue{For any QPU with architecture graph $AG=(V,E)$ and any target QCT algorithm QCT-B, our framework intends to generate an enhanced QCT algorithm that performs better than QCT-B.} This is achieved by employing supervised learning to train a policy ANN which is able to evaluate and recommend possible SWAP operations for input circuits. 
Fig.~\ref{fig:framework} shows the basic modules of the proposed framework and their detailed implementations are elaborated as follows.



{\bf Training Circuits Generation.}
In this module, a large number of training circuits containing $n_l$ layers of gates and $n_q=|V|$ qubits will be randomly generated.
\blue{More precisely, suppose we want to generate $n_c$ circuits for training. Starting from an empty circuit $C$ with $n_q$ qubits, we keep adding to $C$ randomly placed CNOT gates until its depth reaches $n_l \cdot n_c$. The final circuit set is then obtained by sequentially slicing $C$ into $n_c$ sub-circuits each with $n_l$ layers.}


{\bf Label Generation.}
\blue{For each training circuit $C_i$ generated in the previous module, we attach a probability distribution $\mathbf{p}_i$ of recommended SWAPs, which is called the \emph{label} of $C_i$ and is calculated by appropriately invoking the feeding QCT algorithm on $C_i$ and extracting a non-negative number for each edge (corresponds to a SWAP operation) in $AG$. The concrete implementation depends on the specific feeding algorithm. Shortly we shall see two examples in Sec.s~\ref{sec:sahs_label}~and~\ref{sec:mcts_label}.}

{\bf ANN Training.}
\blue{With the circuits $C_i$ and labels $\mathbf{p}_i$ generated in the previous modules as the `teacher', we now train a policy ANN which, for any input circuit $C$ with $n_l$ layers and $n_q$ qubits, outputs a discrete probability distribution $\mathbf{p}$, called the \emph{recommendation probability distribution} of $C$ henceforth, representing how strongly the ANN recommends each valid SWAP operation (corresponding to an edge in $AG$).
}

The ANN training process takes the mean squared error (MSE) and Adam \cite{kingma2014adam} as, respectively, the loss function and optimization method.

The input circuits of the policy ANN are encoded as a sequence of symmetric 0-1 matrices $(M^k : 1\leq k\leq n_l)$ of dimension $n_q\times n_q$, where $M^k_{i,j} = 1$ if and only if in the $k$-th layer there exists a CNOT gate acting on the $i$-th and $j$-th qubits (the direction is irrelevant). Obviously, these matrices are all symmetric. In our implementation, these matrices are further flattened and concatenated into a 0-1 vector.

\begin{figure}[tbp]
	\centerline{\includegraphics[width=0.45\textwidth]{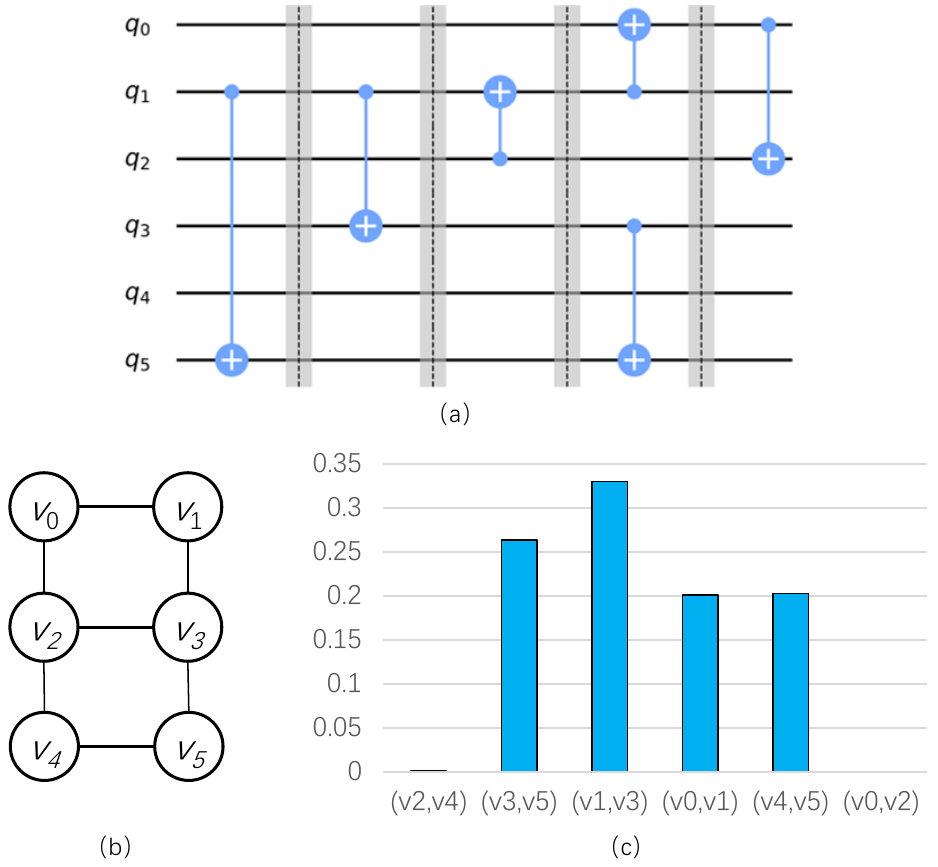}}
	\caption{\blue{An example logical circuit with 6 qubits and 5 layers (a) and the corresponding output probability distribution (c) of the ANN trained under the Grid 2X3 AG (b). The naive mapping, which allocates $q_i$ to $v_i$ for $0\leq i\leq 5$, is used here.} 
	}
	\label{fig:cir_matrix}
\end{figure}

\begin{example}
	Consider the logical circuit and the target $AG$ depicted in Fig.~\ref{fig:cir_matrix} where $n_q = 6$ and $n_l = 5$. Then we have
	\[
	M^1_{1,5} = M^1_{5,1}= \cdots  = M^5_{0,2} = M^5_{2,0} = 1
	\]
	and other entries are all 0.
\end{example}


\blue{For each input circuit $C$, let $\mathbf{p}$ be the recommendation probability distribution of $C$ \blue{output by the ANN}.
Intuitively, the higher the probability a SWAP operation is in $\mathbf{p}$,} the more the ANN `thinks' the corresponding SWAP is promising and the QCT algorithm should be more inclined to select it as the next SWAP in constructing the executable physical circuit.

\begin{example}
	Back to the logical circuit and the target $AG$ depicted in Fig.~\ref{fig:cir_matrix}. \blue{Taking the MCTS algorithm \cite{MCTS_QCT} as the feeding algorithm (cf. Sec.~\ref{sec:mcts})}, the output probability distribution of our trained ANN can be converted to a histogram  that shows to what extent the ANN recommends each valid SWAP \blue{(cf. Fig.~\ref{fig:cir_matrix}c), where the SWAP $(v_1, v_3)$ gets the highest value (around $33\%$). This reflects the fact that if inserting the SWAP $(v_1, v_3)$ then only two SWAPs are required in the whole QCT process (i.e., the SWAP $(v_1, v_3)$ at the beginning and then another SWAP $(v_1, v_3)$ in front of the 4th layer), which is the minimal number we can have.}
	
\end{example}

\blue{
{\bf ANN Embedding.}
The trained ANN can be used in several ways. We can use it to completely replace the evaluation process and thus devise a new algorithm (called ANN-QCT henceforth) for circuit transformation (cf. Sec.~\ref{sec:ANN-QCT}), or use it to assist in the evaluation process when ties need to break (cf. Sec.~\ref{sec:BASE-ANN}), or use it to prune the search tree (cf. Sec.~\ref{sec:sahs} and  Sec.~\ref{sec:mcts}). As shown on  Fig.~\ref{fig:framework} (right), the enhanced algorithm can be obtained by utilizing the trained ANN to modify the evaluation process in the target algorithm. The detailed implementation of the embedding strategy depends on the specific target algorithm used.}



\vspace*{2mm}
\begin{remark}
\blue{The layer number $n_l$ selected in the framework has two direct effects on the ANN Training and Embedding modules. On one hand, a small $n_l$ implies that the circuits generated in the `Training Circuits Generation' module are easy to train; on the other hand, a large $n_l$ may increase the prediction precision of the trained policy ANN. Therefore, we need to trade off easy training with precision by selecting the appropriate value for $n_l$. We refer the reader to Sec.~\ref{sec:sahs_exp} for more detailed discussion.}


\end{remark}

\subsection{ANN-QCT and BASE}
\label{sec:ANN-QCT}
\blue{
As said above, the trained ANN derives a QCT algorithm called ANN-QCT, which, at each step, applies the SWAP operation with the highest recommendation probability provided by the ANN. We now validate the efficacy of the trained ANN by experimentally comparing ANN-QCT with a baseline algorithm (denoted BASE) which utilizes straightforward strategies and considers only the first layer of the current logical circuit.
}



The strategy of selecting the appropriate SWAP at each step plays a key role in  the performance of a QCT algorithm. In BASE, this is achieved with the help of a cost function defined as the total distance of the first layer of CNOT gates:
\begin{equation}\label{eq:totdist}
    \cost{LC,\tau} = \sum\limits_{g \in \mathcal{L}_0(LC)}^{} {\dist_{AG}(g,\tau)}
\end{equation}
where $LC$ is the logical circuit under consideration, $\dist_{AG}(g,\tau)$ is the minimal distance in $AG$ of the two operand physical qubits of CNOT gate $g$ under the mapping $\tau$. Then the first (best) SWAP which minimizes this cost function will be chosen to be added to the physical circuit. The detailed pseudo code can be found in Alg.~\ref{alg:BASE}.

\begin{algorithm}[ht]
	\SetKwData{Left}{left}\SetKwData{This}{this}\SetKwData{Up}{up}
	\SetKwFunction{Union}{Union}\SetKwFunction{FindCompress}{FindCompress}
	\SetKwInOut{Input}{input}\SetKwInOut{Output}{output}
	\Input{An architecture graph $AG=(E,V)$, a logical circuit $LC$, and an initial mapping $\tau_{ini}$.}
	\Output{A physical circuit satisfying the connectivity constraints in $AG$.}
	\caption{BASE}
	\label{alg:BASE}
	\Begin{
	$PC \leftarrow$ all gates in $LC$ executable under $\tau_{ini}$\;
	$LC \leftarrow LC$ with gates in $PC$ deleted\;
    $\tau \leftarrow \tau_{ini}$\;
    
	\While{$LC \neq \emptyset$}
	{
	    $best\_cost \leftarrow \infty$\;
	    \For{all $(v_i,v_j) \in E$}
		{
			$\tau' \leftarrow \tau[\tau^{-1}(v_i)\mapsto v_j, \tau^{-1}(v_j)\mapsto v_i]$\;
			$current\_cost \leftarrow \cost{LC,\tau'}$\;
			\If{$current\_cost < best\_cost$}
    			{$best\_swap \leftarrow \swap{v_i}{v_j}$\;
    			$best\_mapping \leftarrow \tau'$\;
    			$best\_cost \leftarrow current\_cost $\;}
		}
		
	    $\tau \leftarrow best\_mapping$\;
		$C \leftarrow$ the set of all executable gates in $LC$ under $\tau$\;
		$LC \leftarrow LC$ with all gates in $C$ deleted\;
		$PC \leftarrow PC$ by adding $best\_swap$ and all gates in $C$\;
		}
	\Return{$PC$}
	}
\end{algorithm}

\begin{example}
\label{exp:BASE}
Back to the logical circuit, denoted $LC$, in Fig.~\ref{fig:cir_matrix}. Given the naive mapping $\tau_0$ which allocates $q_i$ to $v_i$ for $0\leq i\leq 5$, since the first layer contains only one CNOT gate (involving $q_1$ and $q_5$), it is easy to observe that both $(v_1, v_3)$ and $(v_3, v_5)$ take $\tau_0$ to a new mapping for which the total distance in Eq.~\ref{eq:totdist} achieves its minimum 0. BASE does not distinguish between them and simply chooses the first found one to insert into the output circuit.

\end{example}

We have done experiments on two AGs, Grid 4X4 and IBM Q Tokyo (cf. Fig.~\ref{fig:AG}) for which the ANNs used are trained by SAHS (cf. Sec.~\ref{sec:sahs}) and MCTS (cf. Sec.~\ref{sec:mcts}), respectively. 
The benchmark set consists of 10 circuits each with $|V|$ qubits and 200 randomly placed CNOT gates, where $|V|$ represents the number of vertices in the corresponding AG, i.e., number of physical qubits. 
The results are shown in Table~\ref{tab:benchmark}, and the improvement is calculated as 
\blue{
\begin{align}\label{eq:gate_count_reduction}
    gate\_count\_reduction  = (n_{base} - n_{test})/n_{base}, 
\end{align}
where $n_{test}$ and $n_{base}$ are, respectively, the CNOT overheads brought by the tested algorithm and the baseline algorithm.}  From Table~\ref{tab:benchmark} we can see that, when compared to BASE, both trained ANNs are able to get significantly better solutions (up to 23\%), indicating the accuracy of the outputs of the trained ANN. Surprisingly, ANN-QCT is even better than the Qiskit\footnote{https://qiskit.org/} and comparable to \tket\cite{tket}, which are two widely used industry-level QCT algorithms.

\begin{table}[tbp]
\caption{Experimental results for five QCT algorithms on different AGs. The Comp. column is derived by Eq.~\ref{eq:gate_count_reduction} using BASE as the \blue{baseline} algorithm.}
\label{tab:benchmark}
\begin{tabular}{c|c|c|cc}

AGs                           & Algorithms       & \blue{CNOT Overhead} & Comp.    &  \\
\hline
\multirow{5}{*}{Grid 4X4}    & BASE     & 8388           & 0.00\%  &  \\
                             & Qiskit       & 7590            & 9.51\%  &  \\
                             & \tket       & 5895            & 29.72\% &  \\
                             & ANN-QCT      & 6789            & 19.06\% &  \\
                             & BASE-ANN & 6030            & 28.11\% &  \\
\hline
\multirow{5}{*}{IBM Q Tokyo} & BASE     & 6396            & 0.00\%  &  \\
                             & Qiskit       & 6294            & 1.59\%  &  \\
                             & \tket        & 4584            & 28.33\% &  \\
                             & ANN-QCT      & 4896            & 23.45\% &  \\
                             & BASE-ANN & 4821            & 24.62\% &  \\
\hline
\multirow{4}{*}{Google Sycamore} & BASE     & 22464     & 0.00\%  &  \\
                             & Qiskit       & 17967     & 20.02\%  &  \\
                             & \tket        & 15837     & 29.50\% &  \\
                             & BASE-ANN     & 17904     & 20.30\% &  \\

\end{tabular}
\begin{tablenotes}
\item[1] \blue{Each SWAP gate is decomposed into 3 CNOT gates by default.}
\end{tablenotes} 
\end{table}

\subsection{BASE-ANN}
\label{sec:BASE-ANN}
To exhibit the potential of the ANN Embedding module in our framework, an exemplary \blue{embedding} strategy for BASE is proposed and evaluated.

As shown in Example~\ref{exp:BASE}, it is often the case that more than one SWAP achieve the minimal cost of Eq.~\ref{eq:totdist}, and BASE may choose the wrong one which performs worse in converting the whole circuit. To resolve this problem, we utilize the trained ANN to help further evaluate the best SWAP operations found in the FOR loop of Alg.~\ref{alg:BASE}. 

More specifically, let $best\_swaps$ be the set of SWAPs which achieve the minimal cost with respect to the current logical circuit $LC$ and mapping $\tau$. To break the tie in choosing a best SWAP from $best\_swaps$, the next $n_l$ layers of gates in $LC$ are extracted and a trained policy ANN is invoked to provide a recommendation probability distribution. The SWAP in $best\_swaps$ with the highest recommendation probability will be chosen to be added to the physical circuit. 
We call the enhanced algorithm BASE-ANN henceforth.

\blue{For BASE-ANN, three AGs --- IBM Q Tokyo, Grid 4X4, the 53-qubit Google Sycamore (cf. Fig.~\ref{fig:AG}) --- are tested. The ANNs for two small AGs are trained by SAHS (cf.~Sec.~\ref{sec:sahs}), while the ANN for Sycamore is trained by BASE, which is similar to the one trained by SAHS, except that 
\[PC' \leftarrow \text{SAHS}(LC',AG,\tau',d)\]
in Alg.~\ref{alg:SAHS_label} is replaced with 
\[PC' \leftarrow \text{BASE}(LC',AG,\tau').\]
This is mainly because the label generation process for SAHS and Sycamore is too expensive (in terms of computer time consumption). Besides that,}
the settings and other ANNs used in the experiment are identical to that for ANN-QCT (cf. Table.~\ref{tab:benchmark}).
It can be observed that even this simple embedding strategy suffices to manifest the efficacy of the ANN embedding process (up to 28\% improvement brought by BASE-ANN compared with BASE). \blue{Furthermore, the improvement of BASE-ANN is consistent ($>$20\%) even on Google Sycamore, a QPU with 53 qubits, demonstrating the potential of the proposed method in AGs with large number of qubits.}

\vspace*{2mm}
\blue{In above we have seen that the ANN framework can greatly boost the performance of a baseline QCT algorithm. In the following sections, we shall see that it can also steadily boost the performance of two state-of-the-art QCT algorithms, SAHS\cite{SAHS} and MCTS\cite{MCTS_QCT}.}


\section{Supervised Learning Embedded in SAHS}\label{sec:sahs}
Proposed in \cite{SAHS}, SAHS \blue{(simulated annealing and heuristic search)} is an efficient QCT algorithm which utilizes a double look-ahead mechanism to implement the multi-depth heuristic search. In SAHS,
the search depth, denoted by $d$ henceforth, is a pre-set parameter, through which the trade-off between the running time and the quality, i.e., number of gates, of the transformed circuit can be adjusted. In this section, SAHS is used as \blue{both the feeding and target QCT algorithms} to showcase the efficacy of the proposed framework, and its ANN enhanced counterpart is named SAHS-ANN.

\subsection{Label Generation}
\label{sec:sahs_label}
Described in Alg.~\ref{alg:SAHS_label} is a rough overview for the label generation process based on SAHS, the detailed implementation of which can be found in~\cite{SAHS}. As seen in Alg.~\ref{alg:SAHS_label}, SAHS will be invoked multiple times to evaluate the candidate SWAPs to generate the label for each training circuit. In this label generation process, \blue{the layer number $n_l$ of the training circuits is fixed as 3. Besides, the search depth $d$ for SAHS is set to 2, which is also the default value in its original implementation}.

\begin{algorithm}[ht]
\label{alg:SAHS_label}
	\SetKwData{Left}{left}\SetKwData{This}{this}\SetKwData{Up}{up}
	\SetKwFunction{Union}{Union}\SetKwFunction{FindCompress}{FindCompress}
	\SetKwInOut{Input}{input}\SetKwInOut{Output}{output}
	\Input{An architecture graph $AG$ and a logical circuit $LC$.}
	\Output{A recommendation probability distribution.}
	\caption{Label generation via SAHS}
	
	\Begin{
	$\tau \leftarrow$ the naive mapping\;
	$d \leftarrow 2$\;
    \For{all $e=(v_i,v_j) \in E$}
    {
		$\tau' \leftarrow \tau[\tau^{-1}(v_i)\mapsto v_j, \tau^{-1}(v_j)\mapsto v_i]$\;
		$LC' \leftarrow LC$ with all executable gates under $\tau'$ deleted\;
        $PC' \leftarrow$ SAHS$(LC',AG,\tau',d)$\; 
        \blue{$\mathbf{w}(e) \leftarrow \text{number of SWAPs inserted in }PC'$}\;
    }
    $\mathbf{p} \leftarrow$ the probability distribution proportional to 
    \blue{$\frac{1}{\mathbf{w}(e)+1}$}\;
	\Return{$\mathbf{p}$}
	}
\end{algorithm}
\subsection{ANN Embedding}

In the original SAHS algorithm, the quality of the solution can be significantly improved through increasing $d$. However, this will also exponentially increase the running time, making it unacceptable even for small-size circuits (see data with pruning ratio 0 in Fig.~\ref{fig:sahs_time}). To offset this time overhead, the policy ANN which solely consists of fully connected layers and is trained via \blue{circuits randomly generated in the `Training Circuits Generation' module (cf.~ Sec.~\ref{sec:details})} 
and labels will be embedded in the evaluation process of SAHS.

In SAHS, the QCT problem is reformulated as a search problem, each node and each edge in which contain, respectively, an unfinished physical circuit and a specific SWAP gate to be added to the circuit in its connected parent node. During the search process, the leaf nodes will be opened iteratively until reaching the pre-defined search depth $d$. In the enhanced algorithm SAHS-ANN, the trained policy ANN will be invoked before a node is opened and each candidate SWAP is given a recommendation probability. Then, a proportion, which is  termed as the \emph{pruning ratio} henceforth, of SWAPs will be pruned to decrease the branching factor of the node. Besides that, all other modules are identical to that in SAHS. This ANN-aided pruning mechanism is able to make the search process go deeper while significantly reducing the time overhead when compared to the original SAHS algorithm.

\subsection{Experimental Results}\label{sec:sahs_exp}

\begin{figure}[tbp]
	\centerline{\includegraphics[width=0.3\textwidth]{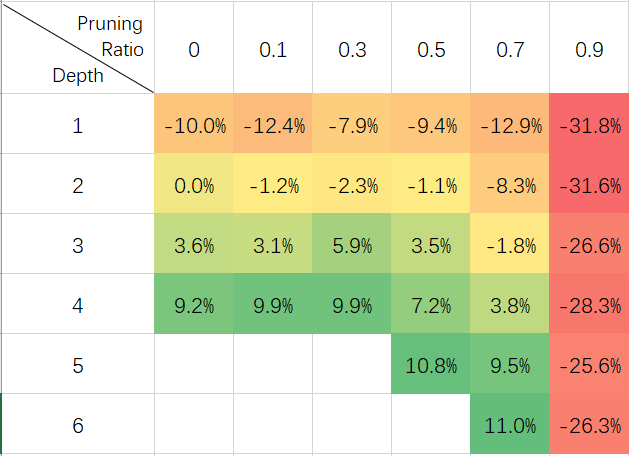}}
	\caption{The improvement of SAHS-ANN vs. SAHS with search depth 2 on Grid 4X4 AG. When the pruning ratio is 0, SAHS-ANN degrades to SAHS.}
	\label{fig:sahs_imp}
\end{figure}

\begin{figure}[tbp]
	\centerline{\includegraphics[width=0.3\textwidth]{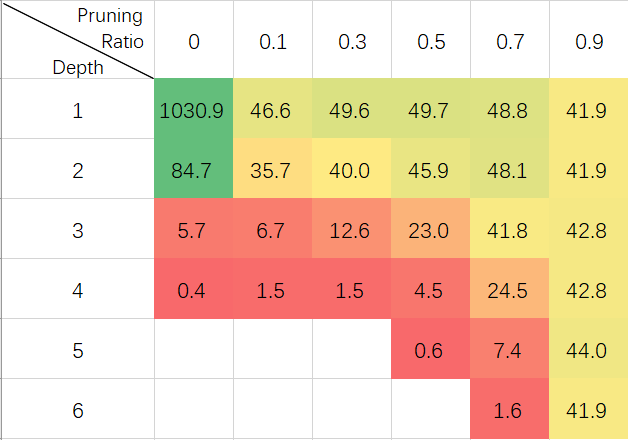}}
	\caption{The time efficiency, which is defined as \blue{the ratio of the number of gates in $LC$ to the running time (seconds),} of SAHS-ANN on Grid 4X4 AG.}
	\label{fig:sahs_time}
\end{figure}

\begin{figure}[tbp]
	\centerline{\includegraphics[width=0.3\textwidth]{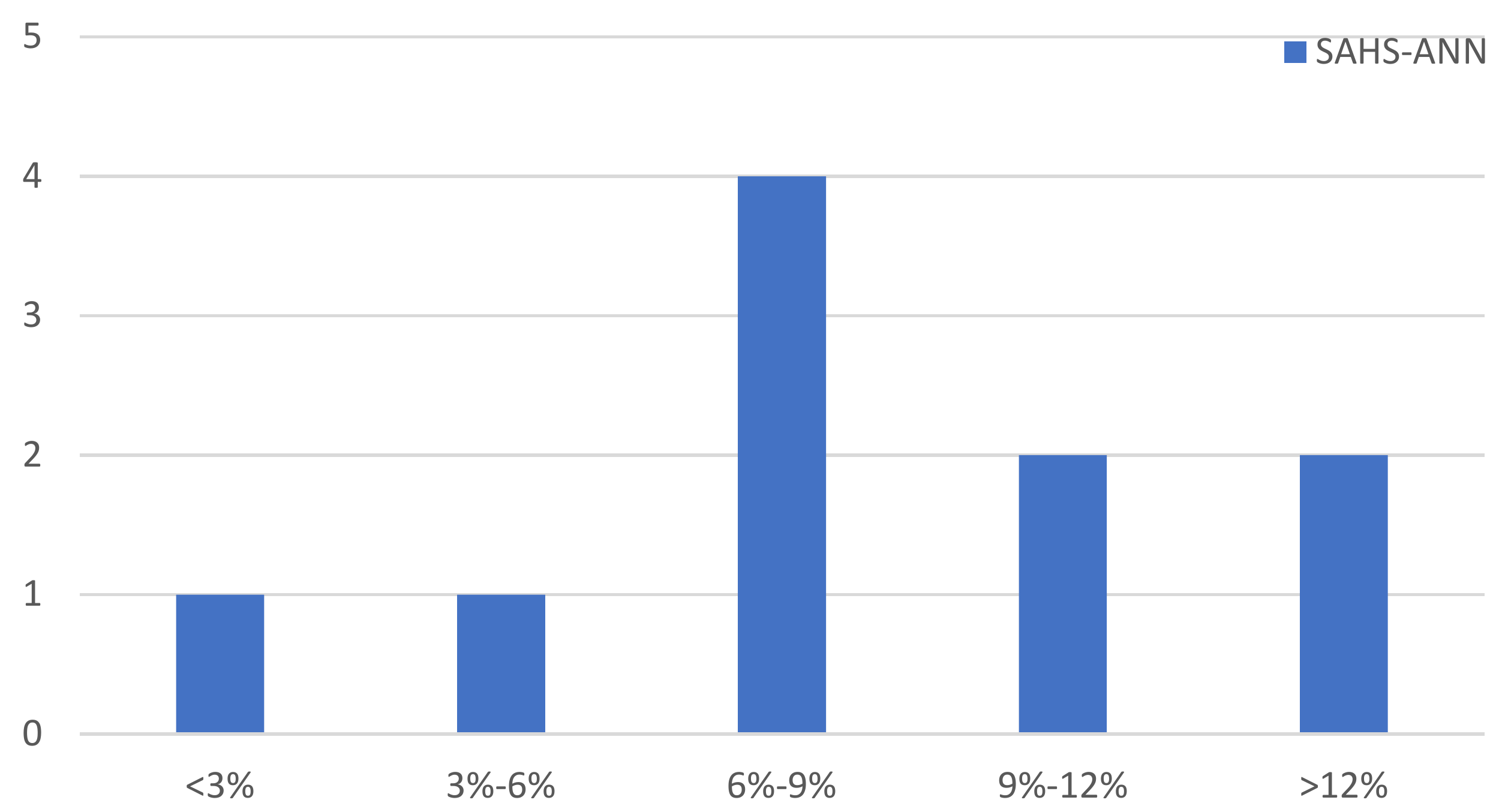}}
	\caption{\blue{The improvement variability of the 10 tested circuits obtained by SAHS-ANN in depth 5 and pruning ratio 0.7, where the vertical (horizontal, resp.) axis represents the number of circuits (the improvement intervals, resp.).}}
	\label{fig:sahs_variability}
\end{figure}

To demonstrate the efficacy of SAHS-ANN, experiments have been done on Grid 4X4.  Fig.s~\ref{fig:sahs_imp} and \ref{fig:sahs_time}  show, respectively, the improvement compared to the original SAHS \blue{with depth 2} and the time efficiency under different depth and pruning ratio settings. \blue{Note that the improvement data in this section are always obtained by comparing with SAHS in depth 2.} 
\blue{From Fig.s~\ref{fig:sahs_imp} and \ref{fig:sahs_time} we observe that, for SAHS,} the quality of solutions can be improved via increasing the search depth (9.2\% when depth is 4)  \blue{at the cost of a dramatic time efficiency degrading} (from 84.7 gates per second to only 0.4). Very attractively, SAHS-ANN is able to obtain a similar quality improvement in depth 5 and pruning ratio 0.7 while its time efficiency is much more promising (7.4 gates per second vs. 0.4) than that of SAHS. \blue{It can be found in Fig.~\ref{fig:sahs_variability} that SAHS-ANN is able to gain more than 6\% improvements on most tested circuits in depth 5 and pruning ratio 0.7, indicating the stability of the proposed algorithm.} Moreover, a 11\% improvement can be derived from SAHS-ANN when the search depth is increased to 6 and, while its time efficiency is still significantly better (1.6 gates per second vs. 0.4).

\blue{It is worth mentioning that, when the pruning ratio reaches 0.9, the performance of SAHS-ANN degrades steeply, making the algorithm almost unusable (cf. Fig.~\ref{fig:sahs_imp}). This is perhaps due to that the ANN used is not always precise and most promising candidates may be pruned away when the ratio is too large.}

\begin{figure}[tbp]
	\centerline{\includegraphics[width=0.3\textwidth]{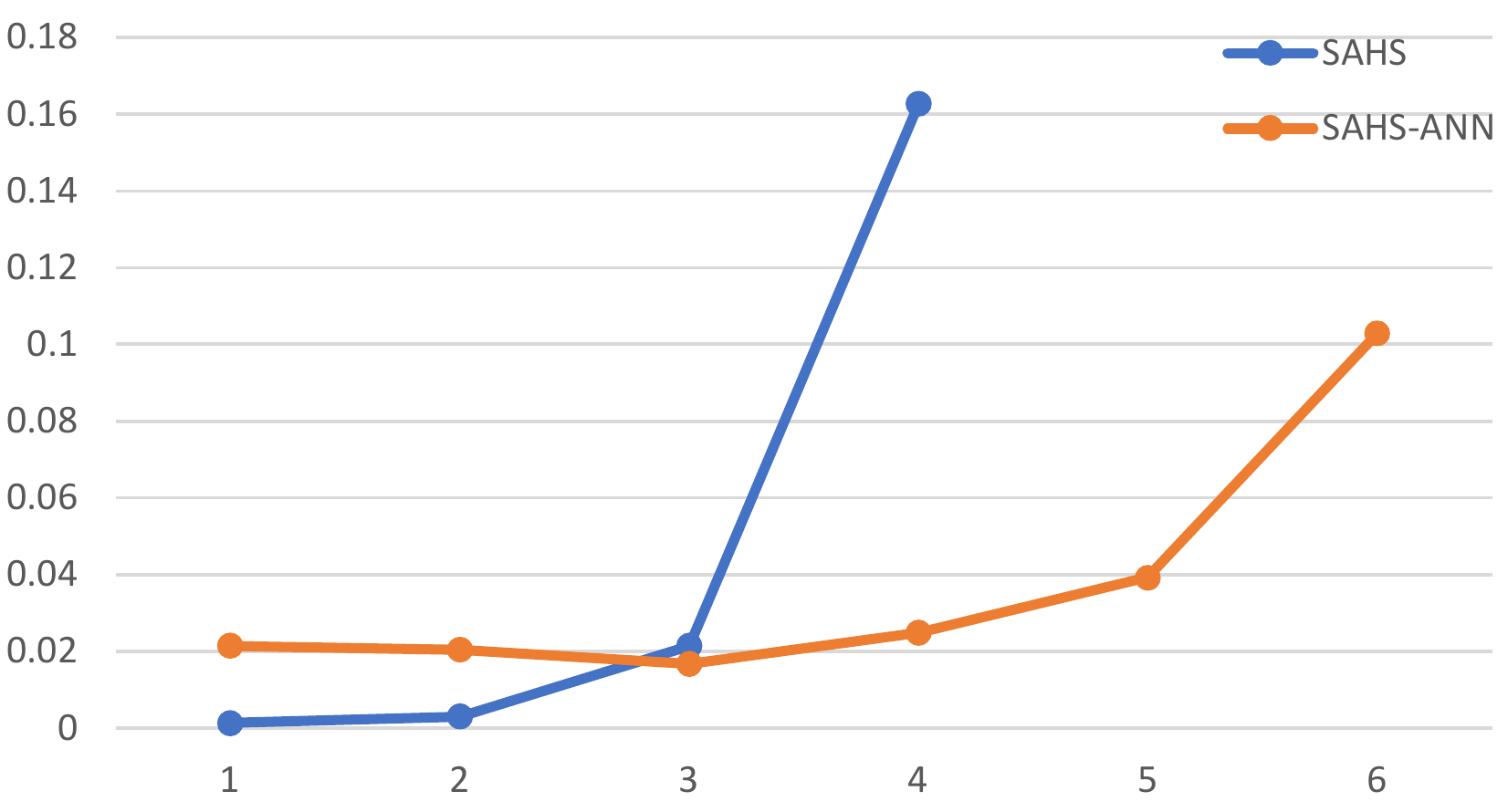}}
	\caption{Running time (seconds per gate in $LC$, vertical axis) of SAHS (blue line) and SAHS-ANN with pruning ratio 0.7 (orange line) vs. search depth (horizontal axis).}
	\label{fig:sahs_depth}
\end{figure}

Now we discuss the influence brought by increasing the search depth to the running time of SAHS and SAHS-ANN. As can be seen from Fig.~\ref{fig:sahs_depth}, the time efficiency of SAHS decreases dramatically as the search process goes deeper, especially when the depth exceeds 3. At the mean time, the time efficiency of SAHS-ANN (with pruning ratio 0.7) is much larger than that of SAHS \blue{(24.5 vs. 0.4 when the search depth is 4)}, which makes it possible for SAHS-ANN to go deeper and, in return, better solutions can be obtained \blue{(e.g., an improvement of 11.0\% with time efficiency 1.6 when the search depth goes to 6)}. 

\begin{figure}[tbp]
	\centerline{\includegraphics[width=0.3\textwidth]{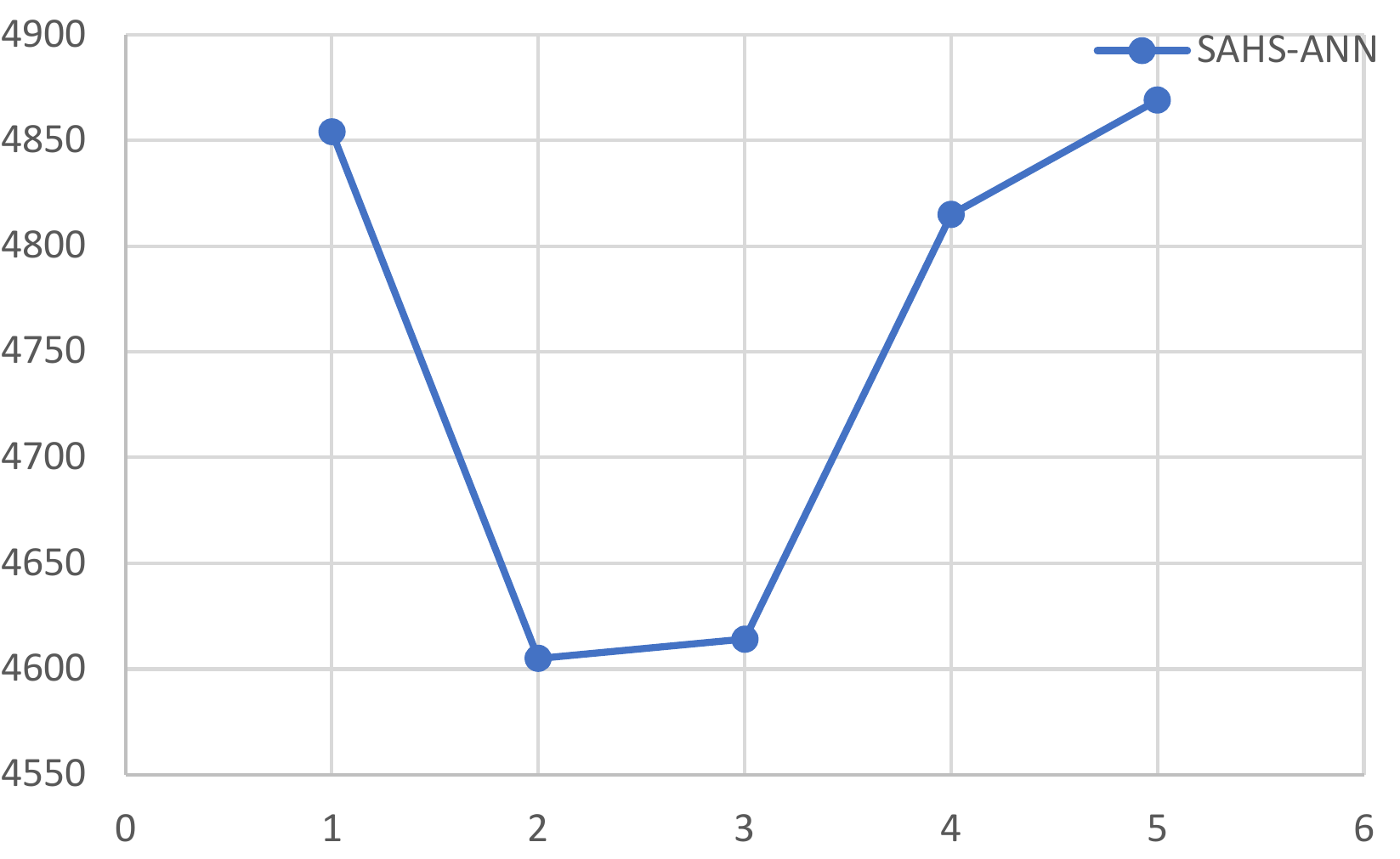}}
	\caption{The \blue{CNOT overhead (vertical axis) in the output circuit brought by} SAHS-ANN with \blue{search} depth 5 and pruning ratio 0.7 vs. various \blue{choices of $n_l$ --- the predefined layer number of training circuits} (horizontal axis).}
	\label{fig:sahs_layer}
\end{figure}


\blue{At last, we evaluate the effect of the parameter $n_l$ (layer number) of the training circuits on Grid 4X4 AG.} Intuitively, the efficacy of embedding an ANN trained with a larger $n_l$ to the target QCT algorithm should be more promising than that with a smaller $n_l$. However, a larger $n_l$ also results in the blow-up of the amount of the information needed to be learned by the ANN, which, in turn, brings a huge challenge for the training process. As depicted in Fig.~\ref{fig:sahs_layer}, better solutions can be derived when the value of $n_l$ is set to 2 or 3, manifesting the rationality of the parameter selection in SAHS-ANN.


\blue{Besides Grid 4X4, experiments are done on IBM Q Guadalupe}\footnote{https://quantum-computing.ibm.com/} with 16 qubits. The results show that the improvement of SAHS-ANN is consistent, 6.0\% in depth 5 and pruning ratio 0.7 and 9.3\% in depth 6 and pruning ratio 0.7.

\blue{To further demonstrate the practicability of the proposed framework, additional experiments are devised on one new benchmark set with 159 realistic circuits,
and Grid 4X4. For this benchmark set, 11 circuits are randomly selected as the test set and the rest are used to compose the training set for the ANN. To make the training set large enough, those circuits are further sliced into multiple sub-circuits each containing $n_l$ layers of CNOT gates.\footnote{Single-qubit gates will be ignored because they have no effect to the QCT process when the objective is minimizing the gate count.}
The results show that a 7.09\% improvement can be obtained in depth 5 and pruning ratio 0.5. 
Furthermore, we also test SAHS-ANN in another benchmark set containing 143 quantum circuits
extracted from the quantum algorithm library in Qiskit. SAHS-ANN works much better in this benchmark set, obtaining a 19.50\% improvement in depth 5 and pruning ratio 0.5 on Grid 4X4.
}

\begin{figure}[tbp]
	\centerline{\includegraphics[width=0.3\textwidth]{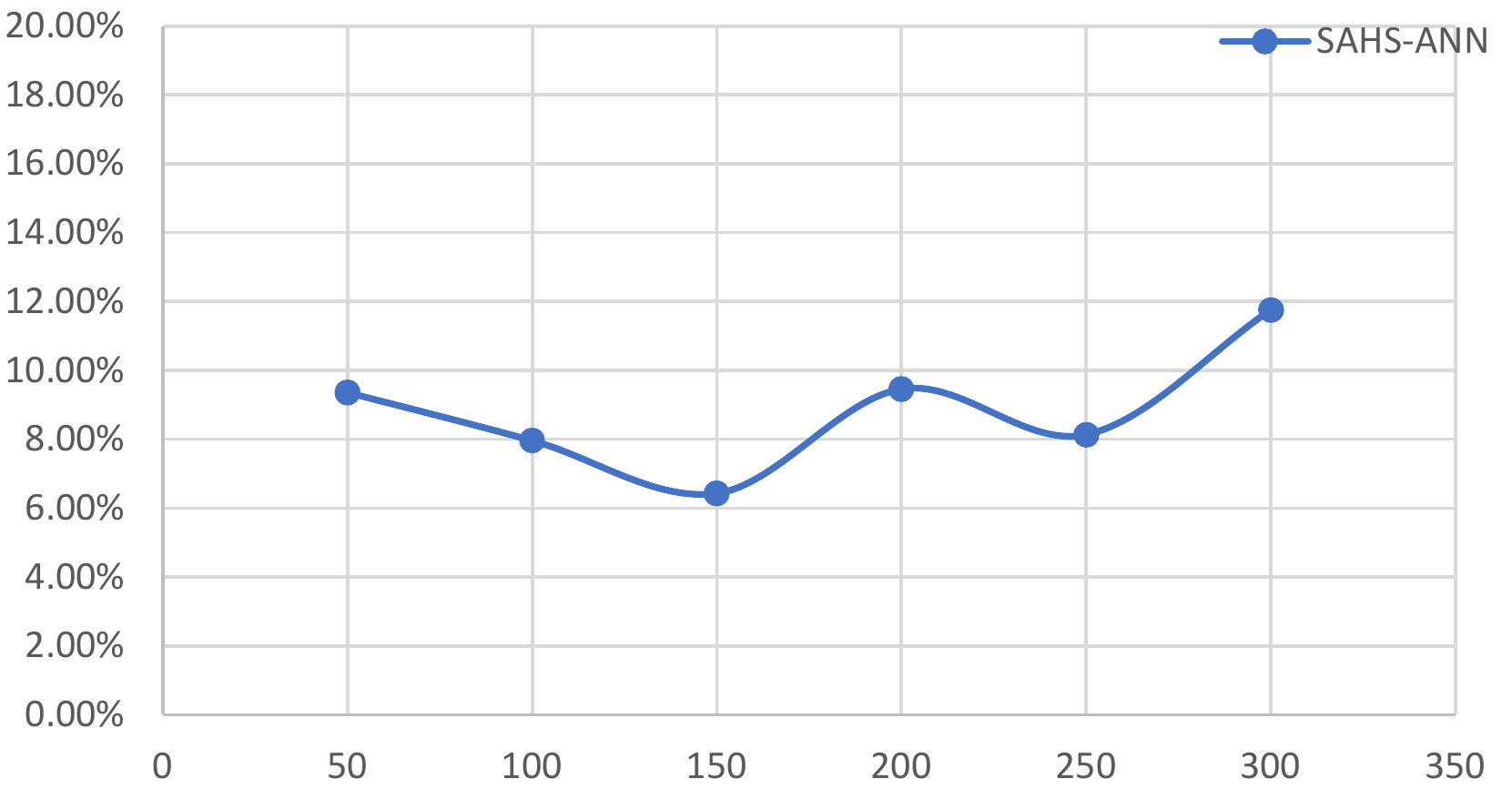}}
	\caption{The average improvement (vertical axis) in the output circuit obtained by SAHS-ANN with search depth 5 and pruning ratio 0.7 vs. various CNOT numbers in the input logical circuits (horizontal axis).}
	\label{fig:sahs_scalability_imp}
\end{figure}

\begin{figure}[tbp]
	\centerline{\includegraphics[width=0.3\textwidth]{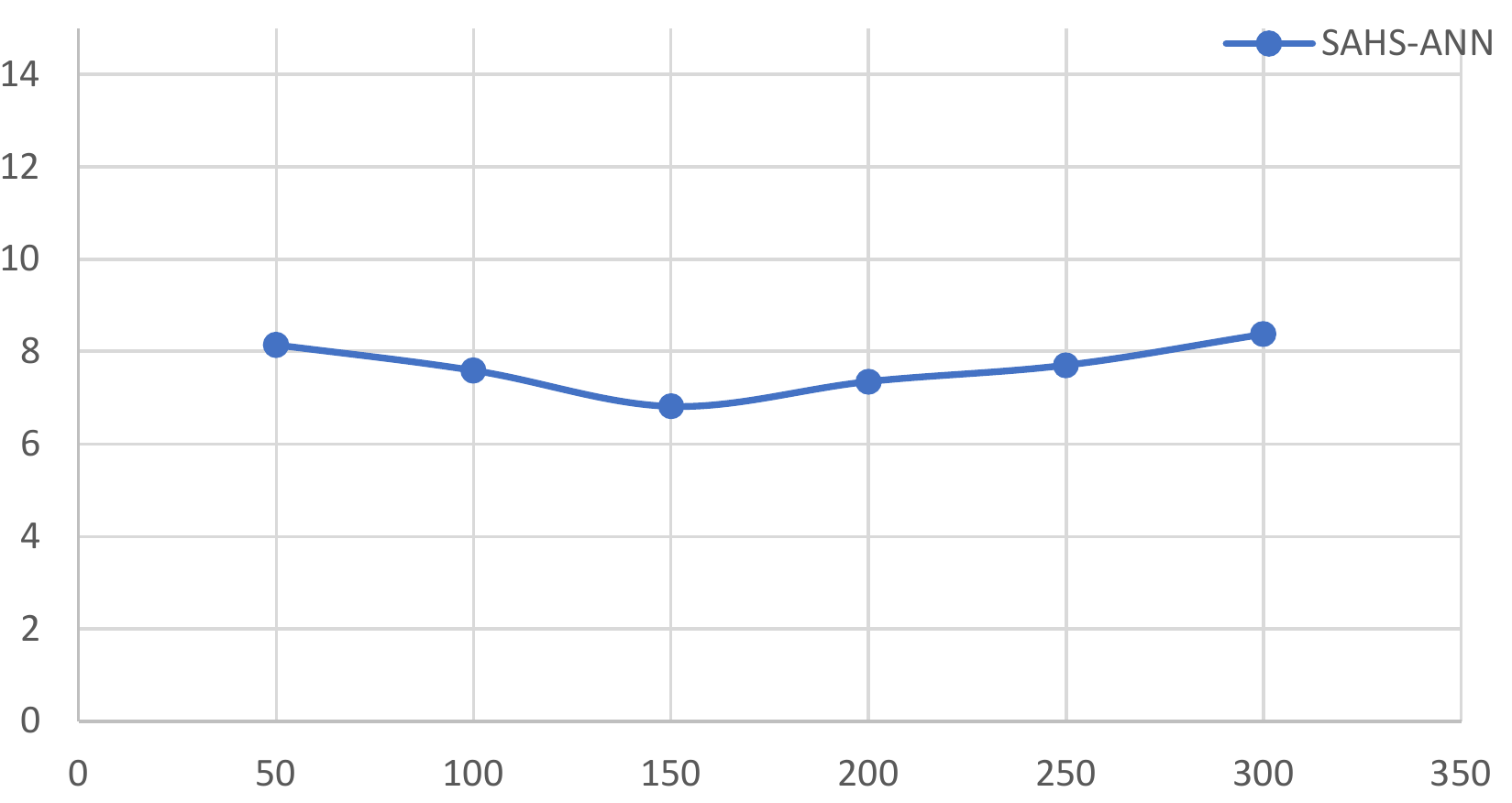}}
	\caption{The average time efficiency (vertical axis) in the output circuit obtained by SAHS-ANN with search depth 5 and pruning ratio 0.7 vs. various CNOT numbers in the input logical circuits (horizontal axis).}
	\label{fig:sahs_scalability_time}
\end{figure}

\blue{To show the scalability of the proposed framework in terms of gate numbers, experiments are also done on a benchmark set consisting of 60 circuits with 16 qubits. Circuits in this benchmark set contain only CNOT gates and their numbers range from 50 to 300. For each number, we transform all circuits under Grid 4X4 using SAHS-ANN with depth 5 and pruning ratio 0.7.  Fig.~\ref{fig:sahs_scalability_imp} shows the average improvement when compared with SAHS with search depth 2.  Fig.~\ref{fig:sahs_scalability_time} shows the average time efficiency, which is the ratio of the number of gates to the running time (seconds). The results show that the improvement and time efficiency of the proposed SAHS-ANN are consistent and steady (ranging from 6.4\% to 12\% and 7 to 8.4 gates per second) under various input circuit sizes, and hence demonstrate the scalability of SAHS-ANN in terms of gate numbers of the input circuits.}

\section{Supervised Learning Embedded in MCTS}\label{sec:mcts}
A Monte-Carlo-Tree-Search based QCT algorithm, abbreviated as MCTS henceforth, is proposed in \cite{MCTS_QCT}.
\blue{MCTS consists of five modules, \emph{Selection}, \emph{Expansion}, \emph{Simulation}, \emph{Backpropagation} and \emph{Decision}. Through each invoking of the \emph{Decision} module, a SWAP gate will be added to the output physical circuit. Before each \emph{Decision}, the \emph{Selection}, \emph{Expansion}, \emph{Simulation} modules will be iteratively executed $n_{bp}$ (a pre-difined parameter) times to provide evaluations for each candidate SWAP gate. Naturally, a larger $n_{bp}$ will increase the precision of the evaluation process at the cost of  significant time overhead.}
MCTS is able to reach a much larger search depth while the complexity is still polynomial and the experimental results show that it exceeds the state-of-the-art QCT algorithms by a large margin in terms of the gate overhead on IBM Q Tokyo.
In this section, \blue{MCTS is used as the feeding and target QCT algorithm} to further demonstrate the efficacy of the proposed framework.

\subsection{Label Generation}
\label{sec:mcts_label}
To label the training circuits, a modified version of MCTS is used to generate the probability distribution of recommended SWAPs (see Alg.~\ref{alg:MCTS_label} for the details).
To increase the reliability of output distributions,
we empirically set the parameter $n_{bp}$ to 200, which is much larger than the original value 20 chosen in \cite{MCTS_QCT}, in the label generation process. Note that the layer number $n_l$ of the training circuits is empirically set to 5.

\begin{algorithm}[ht]
\label{alg:MCTS_label}
	\SetKwData{Left}{left}\SetKwData{This}{this}\SetKwData{Up}{up}
	\SetKwFunction{Union}{Union}\SetKwFunction{FindCompress}{FindCompress}
	\SetKwInOut{Input}{input}\SetKwInOut{Output}{output}
	\Input{An architecture graph $AG$ and a logical circuit $LC$.}
	\Output{A recommendation probability distribution.}
	\caption{Label generation via MCTS}
	
	\Begin{
	$\tau_{ini} \leftarrow$ the naive mapping\;
	$PC \leftarrow$ all gates in $LC$ executable under $\tau_{ini}$\;
	$LC \leftarrow LC$ with gates in $PC$ deleted\;
	$\mathcal{T} \leftarrow$ a search tree with a single (root) node $ (\tau_{ini}, PC, LC)$\;
    \Do{$n_{bp}$ times}
    {
        $\astate \leftarrow$ Select($\mathcal{T}$)\;
        Expand($\mathcal{T},\astate$)\;

		Simulate($\mathcal{T},\astate$)\;
       	Backpropagate($\mathcal{T},\astate$)\;
        
    }
    $\mathbf{p} \leftarrow$ the probability distribution propotional to the scores of all child nodes of root($\mathcal{T}$)\;
	\Return{$\mathbf{p}$}
	}
\end{algorithm}

\subsection{Embedding ANN to MCTS}
As shown in Alg.~\ref{alg:MCTS_label}, the main part of MCTS consists of four modules: \emph{Selection}, \emph{Expansion}, \emph{Simulation}, and \emph{Backpropagation}. Similar to SAHS-ANN, we integrate the trained policy ANN to the \emph{Decision} module to prune the unpromising child nodes of the root. Specifically, when reaching a new root in \emph{Decision}, the ANN is invoked and each child of that root is given a recommendation probability according to its corresponding SWAP operation. Then a proportion, called the \emph{pruning ratio}, of the children are pruned. This ANN-based pruning process helps MCTS to focus only on nodes with more potential. Besides that, all other modules and parameters are identical to that in the original MCTS in \cite{MCTS_QCT}. The ANN-enhanced  MCTS algorithm is called MCTS-ANN henceforth.

\subsection{Experimental Results}\label{sec:mcts_exp}
In this section, experimental results are exhibited to show the performance of MCTS-ANN.

We trained a policy ANN via the strategy introduced in Sec.~\ref{sec:mcts_label} for IBM Q Tokyo (cf. Fig.~\ref{fig:AG})  \blue{with MCTS the feeding QCT algorithm  and $n_l$ being} empirically set to 5. Furthermore, since MCTS is a stochastic process, we run both MCTS and MCTS-ANN  5 times for each circuit and record the minimal gate counts in the output circuits. For running time, the average for each input circuit is recorded.

\begin{figure}[tbp]
	\centerline{\includegraphics[width=0.35\textwidth]{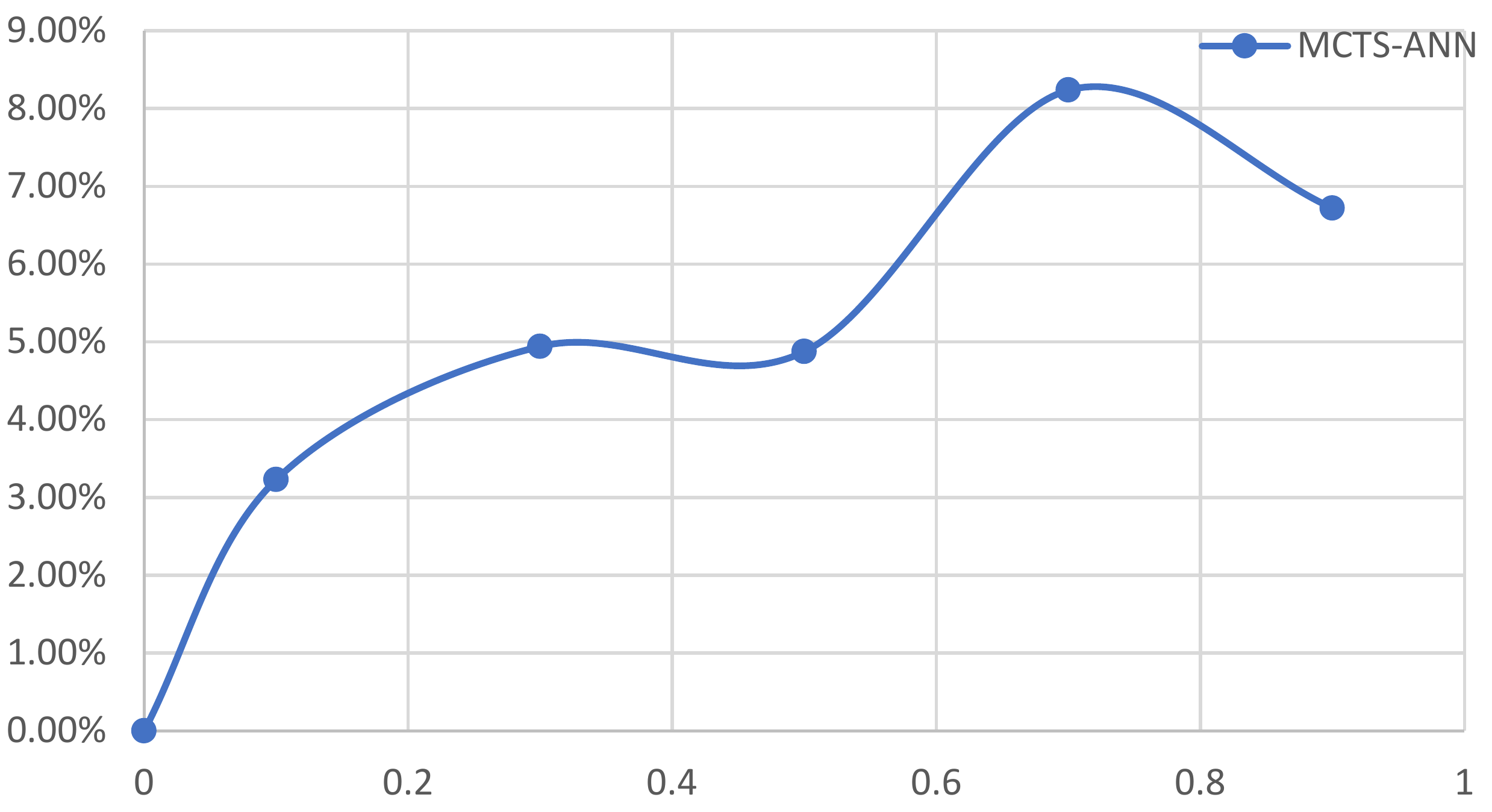}}
	\caption{The performance improvement (cf. Eq.~\ref{eq:gate_count_reduction}) of MCTS-ANN compared with MCTS (vertical axis) vs. various pruning ratios (horizontal axis) on IBM Q Tokyo. When the pruning ratio is 0, MCTS-ANN degrades to MCTS.}
	\label{fig:Q20_gates}
\end{figure}

\begin{figure}[tbp]
	\centerline{\includegraphics[width=0.35\textwidth]{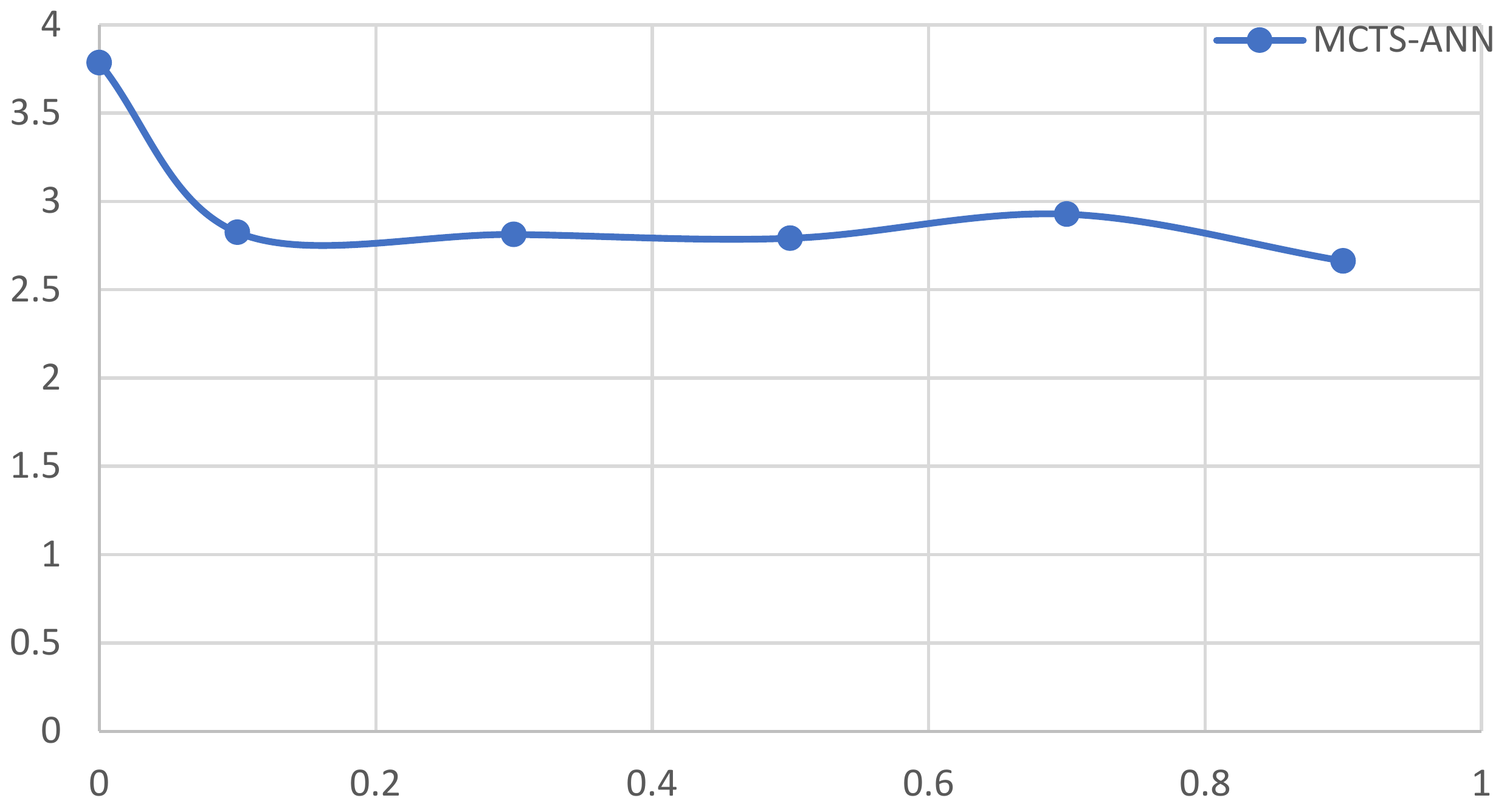}}
	\caption{
		The time efficiency \blue{(i.e., the ratio of the gate number in $LC$ to the running time (seconds))}
		 of MCTS-ANN on IBM Q Tokyo vs. various pruning ratios (horizontal axis). When the pruning ratio is 0, MCTS-ANN degrades to MCTS.}
	\label{fig:Q20_time}
\end{figure}

The performance of MCTS-ANN on IBM Q Tokyo in terms of gate count reduction
and time efficiency are depicted in Figs.~\ref{fig:Q20_gates} and \ref{fig:Q20_time}, respectively, \blue{where the benchmarks used are those used for experiments for Table~\ref{tab:benchmark}.}
\blue{The figures show that the performance of MCTS is effectively improved by MCTS-ANN.}
\blue{For example}, the improvement is the most promising (up to 8\%) when the pruning ratio reaches 0.7. This is because a larger pruning ratio will reduce the branching factor of the root and therefore `encourage' the search process to go deeper under the limited expansion times. As for time efficiency, MCTS-ANN is slightly worse than MCTS, which is acceptable considering the performance improvement (up to 8\%). 

\section{Conclusion} \label{sec:conclusion}
In this paper, we proposed an effective framework based on the idea of using policy ANNs to help existing QCT algorithms in selecting the best 
\blue{SWAP operation}. The policy ANNs can be trained through supervised learning for any specific architecture graph. To demonstrate the effectiveness of the approach, two exemplary ANN-enhanced algorithms, SAHS-ANN and MCTS-ANN, are presented. Experimental results confirm that the trained policy ANN can indeed bring a consistent improvement to their performance on various sets of random and real benchmark circuits and architecture graphs.

\blue{Our current implementation of the framework is far away from being optimal.} As can be seen from Fig.s~\ref{fig:sahs_imp} and \ref{fig:Q20_gates}, the best improvement brought by ANNs is limited (less than or around 10\%). This is possibly due to the poor quality of the currently available training data,
and can be fixed by utilizing more advanced (or exact) algorithms
to generate the labels. One candidate is the recently developed near-exact algorithm TB-OLSQ \cite{tan2020optimal}, which encodes the QCT problem as a 
satisfiability modulo theories optimization problem (SMT) and outperforms several leading heuristic approaches in output circuit quality while being much more efficient than those exact algorithms. 
\blue{Alternatively, labels with better qualities may also be obtained by heuristic algorithms with a radical parameter setting. For example, we can set search depth to 3 or even larger for SAHS where the default value is 2.}

\blue{Both methods are particularly time-demanding for the hardware we used --- a laptop with i7 CPU and 32 GB memory.
For example, the time consumption for generating labels for 1000 random circuits on Grid 4X4 is about 15 minutes using SAHS in depth 2; this figure will be boosted to more than 4 hours when the depth is increased to 3. The situation becomes even worse when the architecture graph has more qubits. In that case, distributed or cloud computing could be used to speed-up the training process and improve the quality of the trained ANN. \blue{Moreover, using the proprietary \tket, instead of SAHS, as the feeding algorithm could reduce the training time by 90\%.} More importantly, this can be done off-line and only one ANN is required for each architecture graph. With affordable computing resource, these approaches are viable and will be one direction of our future research.}


\section*{Acknowledgment}
We thank the reviewers for their very helpful comments and suggestions. 

\bibliographystyle{IEEEtran}
\bibliography{ref}

\appendix
\section{Analysis on scalability issue in label generation}
\blue{We analyze the scalability  in terms of the qubit number both theoretically and experimentally for the label generation process. Assume the target algorithm used is SAHS in search depth 2, the layer number $n_l$ for training circuits is 3, and the AGs are all Grid $k\times k$ like architectures. That is, the AGs have $|V|=k^2$ physical qubits. Recall that the time complexity of SAHS \cite{SAHS} is $O(\ell\cdot |V| \cdot (|E|+|V|/2)^2 \cdot  m)$, where $|E|$ and $m$ are the number of edges in AG and gates in the logical circuit, respectively. Hence, the complexity for generating a label for one circuit is $O(\cdot |V| \cdot (|E|+|V|/2)^2 \cdot m \cdot |E|)$ because at most $|E|$ SWAP gates will be evaluated and for each evaluation process, SAHS has to be invoked for one time.
Then, according to the previously made assumptions, it can be derived that $|E| = 2( {\left| V \right| - \sqrt {\left| V \right|} } )$ and $m \le 1.5|V|$. As a result, the overall complexity for generating one label is bounded by $O( |V|^5 )$.

\begin{figure}[tbp]
	\centerline{\includegraphics[width=0.35\textwidth]{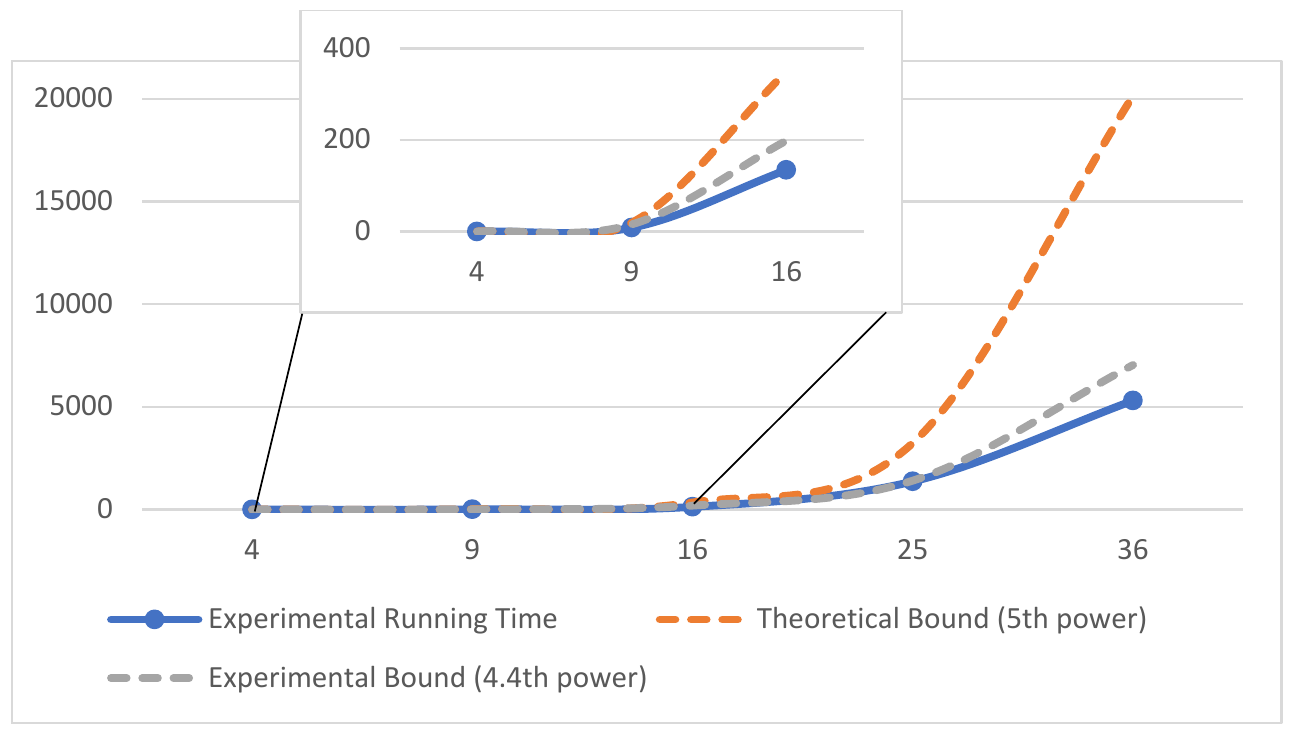}}
	\caption{Running time (seconds, vertical axis) obtained by using SAHS in search depth 2 to generate labels for 100 random circuits vs. various qubit numbers in AGs (horizontal axis), where the subgraph inside is a zooming in to curves between 4 and 16 qubits.}
	\label{fig:scalability_label}
\end{figure}

Experiments are done to further reveal this scalability. For AGs with different qubit numbers, we use SAHS in search depth 2 to generate labels for 100 random circuits each containing 3 layers of CNOT gates, and record the running time. As can be seen from Fig.~\ref{fig:scalability_label}, the real time cost is about the 4.4th power in the number of physical qubits. For example, the time consumption of generating 100 labels via SAHS for Grid 4X4 and Sycamore are about 130 and 45,820 seconds, respectively. Note that $130\times \big(\frac{53}{4\times 4}\big)^{4.4}=25,271$, which is in the same order of magnitude as 45,820.
}


\end{document}